# A Technical Report for Light-Edge: A Lightweight Authentication Protocol for IoT Devices in an Edge-Cloud Environment

**Ali Shahidinejad[1], Mostafa Ghobaei-Arani[1], Alireza Souri[2], Mohammad Shojafar[3] and Saru Kumari[4]**
[1] Department of Computer Engineering, Qom Branch, Islamic Azad University, Qom, Iran
[2] Department of Computer Engineering, Haliç University, Beyoğlu, İstanbul, Turkey
[3] 6GIC/ICS, University of Surrey, Guildford, United Kingdom
[4] Chaudhary Charan Singh University, India

*Abstract*— Selected procedures in [1] and additional simulation results are presented in detail in this report. We first present the IoT device registration in Section I, and we provide the details of fuzzy-based trust computation in Section II. In the end, we show some additional simulation results for formal validation of the Light-Edge under On-the-Fly Model Checker (OFMC) and Constraint-Logic-based ATtack SEarcher (CL-Atse) tools in Section III. See the original paper [1] for more detail.

## I. IoT Device Registration

Table I represents the used variables in this technical report.

TABLE I. Variables and their definitions

| Variable | Definition |
|---|---|
| TC | Trust Center |
| CSP | Cloud service provider |
| $SP_j$ | Service provider j |
| $UID_i$ | ID of device or group i |
| $p_I$ | Password of device i |
| $SID_j$ | ID of server j |
| $b_i$ | Random number chosen by device i |
| $d_j$ | Random number chosen by server j |
| H (.) | Unilateral hash function |
| TS | Time sticker |
| X | Security number of trust center for devices' communications |
| $\oplus$ | XOR operator |
| + | Adjoint operator |
| ΔRT | Threshold of the distance of sending devices' requests to a server |
| ΔT | Threshold of delay |

The first time that device *i* enters the network, it chooses an identifier ($UID_i$) and a password($P_i$). Then using the hash function ($H$), it chooses a random insertion value($b_i$) and encrypts its password and also calculates the value of $A_i$ by Eq. (1).

$$A_i = H(P_i + b_i) \qquad (1)$$

Then it sends a registration message containing ($UID_i . A_i . b_i$) to the trust center. When received, the center calculates $PUID_i$ and $M_i$ by Eq. (2), using variable *X,* which encrypts calculations of the device.

$$PUID_i = h(UID_i + b_i) \qquad (2)$$
$$M_i = h(PUID_i + X)$$

The center then calculates $C_i$ and $D_i$ by Eq. (3)

$$C_i = h(PUID_i + A_i) \qquad (3)$$
$$D_i = M_i \oplus C_i$$

Then, the data of Tables (II) and (III) will be stored in the trust center's database and in the device management compartment, respectively, to get used in any higher-level communications.

TABLE III. Communication data of device i stored in the trust center's database

| $C_i$ | $D_i$ | $b_i$ | h( ) |
|---|---|---|---|

TABLE IIII. Communication data of device i stored in IoT level database

| $UID_i$ | $P_i$ | $D_i$ | $b_i$ | h() |
|---|---|---|---|---|

## II. Trust Computation using a Fuzzy System

We have used Fuzzy Logic Toolbox for developing the fuzzy-based trust computation. The toolbox provides MATLAB functions, Simulink blocks, and graphical tools for designing and analyzing fuzzy-logic-based systems [2]. In the following, we describe the proposed fuzzy-based trust computation module in more detail.

The trust of an IoT device is computed through a fuzzy system by its number of positive and negative actions. A full fuzzy control system is composed of a fuzzifier, an inference mechanism, and a defuzzifier. The fuzzifier module converts control inputs to fuzzy values. A fuzzy variable has values which are defined by linguistic variables (i.e. fuzzy sets or subsets) like low, medium, high, big, great, good, etc. as each of them is defined by a membership function with continuous changes. In the proposed approach, the number of positive and negative actions are the inputs of the system. Through the Mamdani inference mechanism, it is decided that with the input data, what would be the level of trust.

The range for the positive behavior input variable is considered between 0 and 20. As in Fig. 1, for determining this range, the triangle membership function (i.e. TriMF) has been used. The chosen name for each range is assigned by the number of positive behaviors in Table IV.

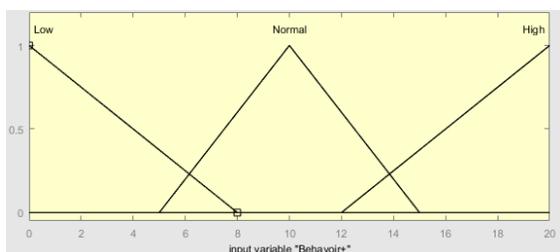

Fig. 1. The fuzzy structure of the device's number of positive behaviors

TABLE IV. Decision structure of input variable by the number of positive behaviors

| Name | Range |
|---|---|
| Low | [-∞ 0 8] |
| Normal | [5 10 15] |
| High | [12 20 +∞] |

The range for the negative behavior input variable is considered between 0 and 5, which is also determined using the TriMF. The name of each range is assigned by the number of negative behaviors in Table V and is illustrated in Fig. 2.

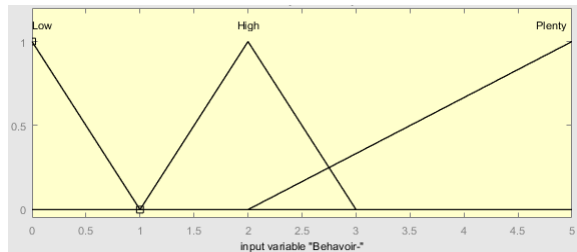

Fig. 2. The fuzzy structure of the device's number of negative behaviors

TABLE V. Decision structure of input variable by the number of negative behaviors

| Name | Range |
|---|---|
| Low | [-1 0 1] |
| High | [1 2 3] |
| Plenty | [2 5 ∞] |

The output variable is the trust degree. Nine groups of trust degree have been considered in the proposed method. According to Fig 3, the triangle membership function has been used for fuzzy variable rules of the trust degree. Each group has been named, and each one's range is explained in Table VI.

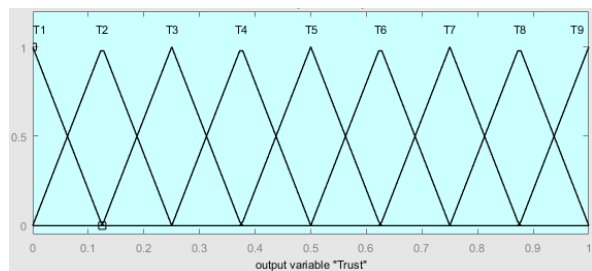

Fig. 3. The fuzzy structure of the output variable of trust degree

TABLE VI. The structure of the output variable of trust degree

| Name | Variable Range |
|---|---|
| T1 | [-∞ 0.00186 0.127] |
| T2 | [0 0.125 0.25] |
| T3 | [0.125 0.25 0.375] |
| T4 | [0.25 0.375 0.5] |
| T5 | [0.375 0.5 0.625] |
| T6 | [0.5 0.625 0.75] |
| T7 | [0.625 0.75 0.875] |
| T8 | [0.75 0.875 1] |
| T9 | [0.875 1 ∞] |

The rule base is a series of fuzzy "if-then" statements that form the heart of the fuzzy inference system. There are two main procedures to determine the fuzzy rules: first, using expert knowledge which is used here, and second, using self-organized educations such as modern algorithms and neural networks. An if-then rule is defined as "if X equals A, then Y equals B" where X and Y are input and output variables and A and B are linguistic values (i.e. membership functions) for these variables. It should be noted that in the Mamdani system, the output is defined in the fuzzy form. The "X equals A" is the assumption, and the "Y equals B" is the result. The calculation rules for measuring the trust degree is defined in Fig. 4.

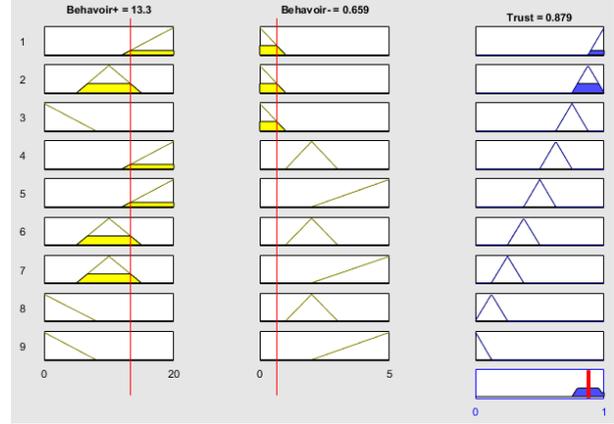

Fig. 5. Defuzzifier structure in the proposed method

```
1. If (Behavoir+ is High) and (Behavoir- is Low) then (Trust is T9) (1)
2. If (Behavoir+ is Normal) and (Behavoir- is Low) then (Trust is T8) (1)
3. If (Behavoir+ is Low) and (Behavoir- is Low) then (Trust is T7) (1)
4. If (Behavoir+ is High) and (Behavoir- is High) then (Trust is T6) (1)
5. If (Behavoir+ is High) and (Behavoir- is Plenty) then (Trust is T5) (1)
6. If (Behavoir+ is Normal) and (Behavoir- is High) then (Trust is T4) (1)
7. If (Behavoir+ is Normal) and (Behavoir- is Plenty) then (Trust is T3) (1)
8. If (Behavoir+ is Low) and (Behavoir- is High) then (Trust is T2) (1)
9. If (Behavoir+ is Low) and (Behavoir- is Plenty) then (Trust is T1) (1)
```

Fig. 4. The structure of Mamdani rules for fuzzy inference

In this section, a series of rules should be mentioned for making decisions. The action of the fuzzy inference motor is similar to human's logic. By executing it on inputs and rules, the output is determined. This is also the way that humans judge. According to input variables, the rules will lead to a fuzzy output production, which then by defuzzification will result in a number for device's priority. The defuzzifier is responsible for the inference. There are different inference methods; for example, the most reliable rule may get chosen as the output. However, it is better to consider an average of all the rules. This method is also called the centroid method. The output is the result of Eq. (4). Figure 5 is an example of the structure of the defuzzifier.

$$Output = \frac{\sum_i x_i \mu(x_i)}{\sum_i \mu(x_i)} \quad (4)$$

Where $\mu(x_i)$ is the returned value of each range of output.

One of the most crucial responsibilities of the trust center is to assess the trust of a device and, using that, determine the device's accessibility level.

By receiving the activity log of a device from the provider, a table similar to Table VII will be created in the trust center's database.

TABLE VII. Device's activity in the trust center

| Action time | $UT_i$ |
|---|---|
| Negative action | NA |
| Positive action | PA |
| Service provider's identifier | $SID_j$ |
| Device's identifier | $UID_i$ |

Through analyzing Table VII, the device's behavior will get evaluated, and its trust degree determined. The trust determination for a device is that the severity facing negative actions is more than the encouragement facing positive ones. For this matter, a time frame structured as Fig. 6 is considered for devices' trust assessment.

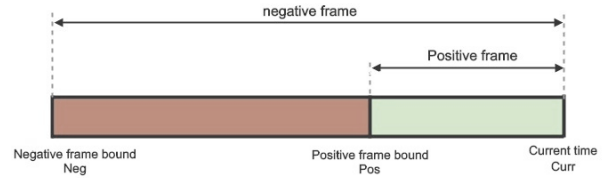

Fig. 6. Time frame structure

The negative and positive bounds of the time frame are defined by Eq. (5) and Eq. (5), respectively.

$$W_p = |Curr - Pos| \quad (5)$$

$$W_n = |Curr - Neg| \quad (6)$$

In Eq. (5) and Eq. (6), Pos and Neg are approvable

time frame boundaries, and Curr represents the current time. According to the table of the device's activity, if positive action is in the $W\_p$ bound, it is considered a positive one. Otherwise, it is disqualified. In the same way, negative actions are important in a bigger time frame; thus, there is more severity evaluating them. By this method, old positive and negative actions that are no longer within these boundaries will be considered obsolete and outdated. Then the current time activity of a device is measured by counting positive and negative actions. This value is positive if positive actions are more frequent than negative ones, and vice versa.

### III. FORMAL VALIDITY OF LIGHT-EDGE

AVISPA software has been used to evaluate the protocol's validation, and MATLAB is used for evaluating the computation time cost and the average communications cost. The On-the-Fly Model Checker (OFMC) and Constraint-Logic-based ATtack SEarcher (CL-Atse) tools are used to formally validate the proposed approach. CL-Atse tool analyzes the security of cryptographic protocols efficiently and versatility [3], and OFMC is a tool that combines two ideas for analyzing security protocols based on lazy, demand-driven search [4].

The proposed approach is compared with three methods Amin et al. [5], Li et al. [6], and Xue et al. [7]. Figures 7-10 illustrate the formal validation of the proposed protocol by OFMC tool in AVISPA with three methods Amin et al. [5], Li et al. [6], and Xue et al. [7].

```
% OFMC
% Version of 2006/02/13
SUMMARY
  SAFE
DETAILS
  BOUNDED_NUMBER_OF_SESSIONS
PROTOCOL
  /home/span/span/testsuite/results/Propsed.if
GOAL
  as_specified
BACKEND
  OFMC
COMMENTS
STATISTICS
  parseTime: 0.00s
  searchTime: 0.11s
  visitedNodes: 3 nodes
  depth: 6 plies
```
Fig. 7. Security evaluation of Light-Edge protocol in examination by OFMC tool

```
% OFMC
% Version of 2006/02/13
SUMMARY
  SAFE
DETAILS
  BOUNDED_NUMBER_OF_SESSIONS
PROTOCOL
  /home/span/span/testsuite/results/Ref1.if
GOAL
  as_specified
BACKEND
  OFMC
COMMENTS
STATISTICS
  parseTime: 0.00s
  searchTime: 0.14s
  visitedNodes: 3 nodes
  depth: 6 plies
```
Fig. 8. Security evaluation of Amin et al. [5] protocol in examination by OFMC tool

```
% OFMC
% Version of 2006/02/13
SUMMARY
  SAFE
DETAILS
  BOUNDED_NUMBER_OF_SESSIONS
PROTOCOL
  /home/span/span/testsuite/results/Ref2.if
GOAL
  as_specified
BACKEND
  OFMC
COMMENTS
STATISTICS
  parseTime: 0.00s
  searchTime: 0.09s
  visitedNodes: 6 nodes
  depth: 9 plies
```
Fig. 9. Security evaluation of Li et al. [6] protocol in examination by OFMC tool

```
% OFMC
% Version of 2006/02/13
SUMMARY
  UNSAFE
DETAILS
  BOUNDED_NUMBER_OF_SESSIONS
PROTOCOL
  /home/span/span/testsuite/results/Ref2.if
GOAL
  as_specified
BACKEND
  OFMC
COMMENTS
STATISTICS
  parseTime: 0.00s
  searchTime: 0.07s
  visitedNodes: 3 nodes
  depth: 5 plies
```
Fig. 10. Security evaluation of Xue et al. [7] protocol in examination by OFMC tool

Figures 11-14 illustrate the formal validation of the proposed protocol by CL-AtSe tool in AVISPA with three methods Amin et al. [5], Li et al. [6], and Xue et al. [7].

```
SUMMARY
 SAFE
DETAILS
 BOUNDED_NUMBER_OF_SESSIONS
 TYPED_MODEL
PROTOCOL
 /home/span/span/testsuite/results/Proposed.if
GOAL
 As Specified
BACKEND
 CL-AtSe
STATISTICS
 Analysed   : 6 states
 Reachable  : 6 states
 Translation: 0.08 seconds
 Computation: 0.03 seconds
```

Fig. 11. Security evaluation of Light-Edge protocol in examination by CL-AtSe tool

```
SUMMARY
 SAFE
DETAILS
 BOUNDED_NUMBER_OF_SESSIONS
 TYPED_MODEL
PROTOCOL
 /home/span/span/testsuite/results/Proposed.if
GOAL
 As Specified
BACKEND
 CL-AtSe
STATISTICS
 Analysed   : 6 states
 Reachable  : 6 states
 Translation: 0.08 seconds
 Computation: 0.03 seconds
```

Fig. 12. Security evaluation of Amin et al. [5] protocol in examination by CL-AtSe tool

```
SUMMARY
 UNSAFE
DETAILS
 BOUNDED_NUMBER_OF_SESSIONS
 TYPED_MODEL
PROTOCOL
 /home/span/span/testsuite/results/Ref2.if
GOAL
 As Specified
BACKEND
 CL-AtSe
STATISTICS
 Analysed   : 4 states
 Reachable  : 3 states
 Translation: 0.05 seconds
 Computation: 0.02 seconds
```

Fig. 13. Security evaluation of Li et al. [6] protocol in examination by CL-AtSe tool

```
SUMMARY
 SAFE
DETAILS
 BOUNDED_NUMBER_OF_SESSIONS
 TYPED_MODEL
PROTOCOL
 /home/span/span/testsuite/results/Ref3.if
GOAL
 As Specified
BACKEND
 CL-AtSe
STATISTICS
 Analysed   : 5 states
 Reachable  : 5 states
 Translation: 0.06 seconds
 Computation: 0.01 seconds
```

Fig. 14. Security evaluation of Xue et al. [7] protocol in examination by CL-AtSe tool

We can conclude that the Light-Edge and Amin et al. [5] are safe under both OFMC and CL-Atse tests, while Li et al. [6] approach is unsafe under the CL-Atse test, and Xue et al. [7] is unsafe under the OFMC test.

**Ali Shahidinejad** is an Assistant Professor at Qom Branch, Islamic Azad University. Contact him at a.shahidinejad@qom-iau.ac.ir.

**Mostafa Ghobaei-Arani** is an Assistant Professor at Qom Branch, Islamic Azad University. Contact him at m.ghobaei@qom-iau.ac.ir.

**Alireza Souri** is an Assistant Professor at Department of Computer Engineering, Haliç University, İstanbul, Turkey. Contact him at alirezasouri@halic.edu.tr.

**Mohammad Shojafar** is an Associate Professor at 6G innovation Centre (6GIC), University of Surrey, Guildford, United Kingdom. Contact him at m.shojafar@surrey.ac.uk.

**Saru Kumari** is an Assistant Professor at Department of Statistics, Chaudhary Charan Singh University, Meerut, India. Contact her at saryusiirohi@gmail.com.